\documentclass{llncs}

\usepackage{amsmath}
\usepackage{graphicx}
\usepackage{subfigure}

\usepackage{lineno}
\newcommand{\iseij}[6]{\left\{
    \begin{array}{ccc}
      {#1}&{#2}&{#3}\\
      {#4}&{#5}&{#6}
    \end{array}\right\}}

\newcommand{\itrej}[6]{\left(
    \begin{array}{ccc}
      {#1}&{#2}&{#3}\\
      {#4}&{#5}&{#6}
    \end{array}\right)}

\title{The screen representation of vector coupling coefficients
       or Wigner $3j$ symbols: exact computation and illustration
       of the asymptotic behavior}
\titlerunning{Screen representation of $3j$ symbols}

\author{Ana Carla P. Bitencourt\inst{1} \and Mirco Ragni\inst{1}
       \and Robert G. Littlejohn\inst{2} \and Roger Anderson\inst{3}
       \and Vincenzo Aquilanti\inst{4,}\inst{5}}

\authorrunning{Ana Carla P. Bitencourt}

\institute{Departamento de F\'isica, Universidade Estadual de Feira de Santana, Brazil \email{ana.bitencourt@gmail.com} \and
           Department of Physics, University of California, Berkeley, CA 94720, U.S.A. \and
           Department of Chemistry, University of California, Santa Cruz, CA 95064, U.S.A. \and
           Dipartimento di Chimica, Biologia e Biotecnologie, Universit\`a di Perugia, Italy, 06123 \and
           Instituto de F\'isica, Universidade Federal da Bahia, Brasil}

\begin{document}

\maketitle

\begin{abstract}
The Wigner $3j$ symbols of the quantum angular momentum theory are
related to the vector coupling or Clebsch-Gordan coefficients and
to the Hahn and dual Hahn polynomials of the discrete orthogonal
hyperspherical family, of use in discretization approximations. We
point out the important role of the Regge symmetries for defining
the screen where images of the coefficients are projected, and for
discussing their asymptotic properties and semiclassical behavior.
Recursion relationships are formulated as eigenvalue equations,
and exploited both for computational purposes and for physical
interpretations.
\end{abstract}


\section{Introduction}\label{s1}

We consider here the important coefficients which describe vector
couplings in quantum mechanics. For an introduction, relevant to
the great variety of applications in chemistry and physics, see
Ref. \cite{zare}. They are known as Clebsch-Gordan coefficients
and also as Wigner's $3j$ symbols, and mathematically are related
to Hahn and dual Hahn-polynomials \cite{avcf1995}. For their
general properties, specifically from the viewpoint of asymptotic
and semiclassical analysis, see Ref.
\cite{aquilanti2007semiclassical} and references therein. They
stand among the simplest spin networks and from a modern viewpoint
many of their properties can be derived from those of Wigner's
$6j$ symbols (or Racah's coefficients). Therefore this work can be
considered as a continuation of a series of previous papers in
these Lecture Notes
\cite{lncs1.2013,lncs2.2013,bitencourt2012exact}. For relevant
exact and semiclassical approaches, see Ref.
\cite{miller1974,schgora,schgorb}; Ref. \cite{sprung2009}
illustrates some details of specific features.

The next section discusses a key property, the Regge symmetries,
crucial to our treatment and neglected in most of previous work.
We exploit it to define the screen for producing images of their
values and features (Sec. \ref{s3}). Sec. \ref{s4} reports the
basic recurrence relationships explicitly as a set of two dual
eigenvalue equations. Detailed derivations and main implications
will not be given, since a main focus of this presentation is an
account of caustics and ridges (Sec. \ref{s5}) limiting the
classical - quantum boundaries and in general the scenario for the
illustrations (Sec. \ref{s6}). Conclusions and final remarks are
in Sec. \ref{s7}. An Appendix lists permutational and mirror
symmetries, which are referred to in the main text.

\section{Regge symmetries}\label{s2}

We need to establish notations and conventions to be exploited in
the definition of the screen for the representation of vector
coupling coefficients or Wigner's $3j$ symbols. They are related
by \cite{zare}
\begin{equation}\label{eq1}
\langle a ~ \alpha, ~ b ~ \beta ~ |x ~ -\gamma\rangle =
(-1)^{a-b-\gamma}(2x+1)^{1/2}
\itrej{a}{b}{x}{\alpha}{\beta}{\gamma}
\end{equation}
$$(\alpha+\beta+\gamma = 0).$$
In the following, we will consider Regge symmetries
\cite{ponzano1968semiclassical,mohanty2003regge,roberts1999classical,aquilanti2013volume};
they play an important role in our treatment and are much less
evident than the usual permutational ones (see Appendix). We will
often be guided by analogy with the Racah's recoupling
coefficients or Wigner's $6j$ symbols
\cite{lncs1.2013,lncs2.2013,neville1971technique,robyu}, from
which the $3j$ symbols can be connected through a limiting
procedure, given here with no specification of phases or
normalizations
\cite{ponzano1968semiclassical,biedenharnlouckch5.8,rbfaal.10}
\begin{equation}\label{eq2}
 \iseij{a}{b}{x}{C}{D}{Y} \longrightarrow
 \itrej{a}{b}{x}{D-Y}{Y-C}{C-D}.
\end{equation}
Here, capital letters denote very large entries (specifically,
magnitudes larger than one order with respect to $\hbar$), and the
arrow indicates that their limit at infinity is taken. Comparison
with (\ref{eq1}) gives
\begin{eqnarray}\label{eq3}
 D-Y\longrightarrow \alpha, \quad Y-C\longrightarrow \beta, \quad C-D\longrightarrow
 \gamma~,
\end{eqnarray}
with $\gamma=-\alpha-\beta$.

From Ref. \cite{varsh}, p. 298, Eq. (3), fourth equality, we get
\begin{eqnarray}
\iseij{(a+x-C+Y)/2}{~(a+b+C-D)/2}{~(b-x+D-Y)/2}{~(-a+x+C+Y)/2}{~(a-b+C+D)/2}{~(b-x+D+Y)/2}\nonumber
\\\longrightarrow
\itrej{(a+x+\beta)/2}{~(a+b+\gamma)/2}{~(b-x+\alpha)/2}{-b+(a+x-\beta)/2}{~-x+(a+b-\gamma)/2}{~-a+(b+x-\alpha)/2}\label{varsh1}
\end{eqnarray}
and from the third equality we obtain
\begin{eqnarray}
\iseij{(-a+b+C+D)/2}{~(a-b+C+D)/2}{~x}{(a+b-C+D)/2}{~(a+b+C-D)/2}{~Y}\label{varsh2.1}\\
=
\iseij{(a+b-C+D)/2}{~(a+b+C-D)/2}{~x}{(-a+b+C+D)/2}{~(a-b+C+D)/2}{~Y}\label{varsh2.2}\\
\longrightarrow
\itrej{(a+b-\gamma)/2}{~(a+b+\gamma)/2}{~x}{(a-b-\beta+\alpha)/2}{~(a-b+\beta-\alpha)/2}{~-a+b}.\label{varsh2.3}
\end{eqnarray}
Eq. (\ref{varsh2.2}) follows from (\ref{varsh2.1}) by a
permutational symmetry (see Appendix). In this way, from the Regge
symmetries for the $6j$ symbol, we obtain both the two Regge
symmetries of the $3j$ symbol (Ref. \cite{varsh}, p. 245, Eq.
(9)).

Crucial to this paper will be the second relationship, Eqs.
(\ref{varsh2.2})-(\ref{varsh2.3}). It is convenient to change
variables \cite{lncs2.2013,neville2006volume1,neville2006volume2}.
Defining
\begin{eqnarray}
 \delta=\frac{\alpha-\beta}{2}, \quad
 \sigma=\frac{-\gamma}{2}=\frac{\alpha+\beta}{2}~,
\end{eqnarray}
we obtain
\begin{eqnarray}
 \alpha=\sigma+\delta, \quad \beta=\sigma-\delta~,
\end{eqnarray}
and
\begin{equation}\label{varsh2}
 \itrej{a}{b}{x}{\sigma+\delta}{~\sigma-\delta}{~\gamma} = \itrej{(a+b)/2+\sigma}{~(a+b)/2-\sigma}{~x}{(a-b)/2+\delta}{~(a-b)/2-\delta}{~-a+b}
 \equiv \itrej{a'}{b'}{x'}{\alpha'}{\beta'}{\gamma'}.
\end{equation}
Therefore
\begin{eqnarray}
x'= x, \quad \delta'\equiv\frac{\alpha'-\beta'}{2} = \delta, \quad
\sigma'\equiv\frac{\alpha'+\beta'}{2} = \frac{a-b}{2}~,
\end{eqnarray}
where ``$\equiv$'' indicates the introduction of new symbols,
establishing here the correspondence between $3j$ symbols which
are identical by Regge symmetry, and denoted by unprimed and
primed entries. Note invariance of $x$ and $\delta$ with respect
to the Regge symmetry: we will exploit this next in the definition
of the screen.

\section{The screen}\label{s3}
The allowed values of $x=x'$ can be obtained from the triangular
relationship among $a$, $b$, and $c$, $|a-b|\leq x\leq a+b$, etc
and the limitation of projections by $|\alpha|\leq a$,
$|\beta|\leq b$:
\begin{equation}\label{range:x1}
 \textrm{max}(|a-b|,|\alpha+\beta|) \leq x \leq a+b~,
\end{equation}
so the range of $x$, namely $[x_{max}-x_{min}+1]$, is the smallest
of four numbers:
\begin{eqnarray}
&a+b+a-b+1=2a+1&\label{range:x2.1}\\
&a+b-a+b+1=2b+1&\\
&a+b-\alpha-\beta+1=a+b+2\sigma+1&\\
&a+b+\alpha+\beta+1=a+b-2\sigma+1&~.\label{range:x2.4}
\end{eqnarray}
We will now show that
\begin{equation}\label{range:delta1}
\textrm{max}(-a-\sigma,-b+\sigma) \leq \delta \leq \textrm{min}
(a-\sigma,b+\sigma)~.
\end{equation}

In fact, being $\alpha$ and $\beta$ projections of $a$ and $b$
respectively, we have
\begin{eqnarray}\label{range1}
&-a\leq \alpha =\sigma+\delta \leq +a&\\
&-\alpha-\sigma \leq  \delta \leq a-\sigma&
\end{eqnarray}
and
\begin{eqnarray}\label{range2}
&-b\leq \beta  =\sigma-\delta \leq +b& \\
&-b-\sigma      \leq -\delta \leq b-\sigma& \\
&-b+\sigma \leq \delta \leq b+\sigma
\end{eqnarray}
proving Eq. (\ref{range:delta1}).

Therefore the range of $\delta$ is the minimum of the four
numbers:
\begin{eqnarray}
&a-\sigma+a+\sigma +1= 2a+1&\label{range:delta2.1}\\
&a+b-2\sigma+1&\\
&a+b+2\sigma+1&\\
&b+\sigma+b-\sigma +1=2b+1&\label{range:delta2.4}
\end{eqnarray}
(to be compared with Eq. (\ref{range:x2.1})-(\ref{range:x2.4})).
Being the range of $\delta$ = range of $x$, any plot having $x$
and $\delta$ as Cartesian axes is a square screen.

As in Ref. \cite{lncs1.2013} for the $6j$s, we recognize the
surprising manifestation of the Regge symmetry in both Eqs.
(\ref{range:x2.1})-(\ref{range:x2.4}) and
(\ref{range:delta2.1})-(\ref{range:delta2.4}): let's rewrite
compactly the relationships between conjugates
\begin{center}
\begin{tabular}{lll}
$a' = \frac{a+b}{2}+\sigma$ & ~~~~~~~ & $a = \frac{a'+b'}{2}+\sigma'$ \\
$b' = \frac{a+b}{2}-\sigma$ & ~~~~~~~ & $b = \frac{a'+b'}{2}-\sigma'$ \\
$\sigma' =  \frac{a-b}{2}$  & ~~~~~~~ & $\sigma = \frac{a'-b'}{2}$ \\
\end{tabular}
\end{center}
These permit to establish the convention of electing to refer to
one of the Regge conjugates which contains the minimum of the four
quantities in these equations, and to identify it with $a$,
possibly by a permutational symmetry (see Appendix). The screen
will therefore be $(2a+1) \times (2a+1)$.

\section{Recurrence relationships as eigenvalue
equations}\label{s4}
We can now write the two basic three-term recurrence relationships
for $3j$ coefficients, modifying them to appear as symmetric
eigenvalue equations. From equations (9a), (9b), and (9c) of
Ref.\cite{schgora}, identifying
\begin{equation}\label{eq:recur1}
\itrej{j_1}{j_2}{j_3}{m_1}{m_2}{m_3} \equiv
\itrej{x}{a}{b}{~-2\sigma}{~\sigma+\delta}{~\sigma-\delta} =
\itrej{a}{b}{x}{~\sigma+\delta}{~\sigma-\delta}{~-2\sigma}
\end{equation}
\begin{eqnarray}
m_2=\sigma+\delta, \quad
m_3=\sigma-\delta, \quad
m_2m_3=(\sigma^2-\delta^2)
\end{eqnarray}
one obtains the recurrence equation in the variable $\delta$, with
a range, according to the convention of the previous section, from
$a-\sigma$ to $a+\sigma$.

We find it convenient to write the three-term relationship in terms of the orthonormal functions:
\begin{equation}
U_{ab\sigma}(x,\delta) \equiv \sqrt{2x+1}
\itrej{a}{b}{x}{~\sigma+\delta}{~\sigma-\delta}{~-2\sigma}~.
\end{equation}
This notation simplifies the recurrence relations and makes it
clear that the ``screen'' for $3j$ depends on the three
parameters: $a$, $b$, and $\sigma$, to be compared to the screen
for the $6j$ case which depends on the four parameters denoted
$a$, $b$, $c$, and $d$ in Ref. \cite{lncs1.2013,lncs2.2013}.
\begin{equation}\label{eq:recur2}
p(\delta+1)U_{ab\sigma}(x,\delta+1) +
p_0(\delta)U_{ab\sigma}(x,\delta)
+p(\delta)U_{ab\sigma}(x,\delta-1)=0~,
\end{equation}
where
\begin{eqnarray}\label{eq:recur2.1}
p(\delta)& = &\left[(a-\sigma-\delta-1)(a+\sigma+\delta)(b+\sigma-\delta+1)(b-\sigma+\delta)\right]^{1/2}\\
p_0(\delta)& = &a(a+1) + b(b+1) - x(x+1) + 2(\sigma^2-\delta^2)~.
\end{eqnarray}
Therefore the three - term recursion Eq. (\ref{eq:recur2}) can be
viewed as an eigenvalue equation, where $\lambda = x(x+1) - a(a+1)
- b(b+1)$ are the eigenvalues. This relationship can be related to
the definition of Hahn polynomials \cite{avcf1995,fcv.03},
relevant members of the class of discrete hypergeometric
polynomial families.

The dual three-term recursion equation is in the variable $x$ and
is obtained explicitly, again from Ref. \cite{schgora} Eqs. (6a),
(6b) and (6c), through symmetrization and the normalization by
$(2x+1)^{1/2}$. The normalization plays the same role as in our
treatment of the recurrence for the $6j$ symbol in
\cite{lncs1.2013}; we obtain the recurrence in $x$:
\begin{equation}\label{eq:recur3}
q(x+1)U_{ab\sigma}(x+1,\delta) + q_0(x)U_{ab\sigma}(x,\delta)
+q(x)U_{ab\sigma}(x-1,\delta) = 0~,
\end{equation}
where
\begin{equation}\label{eq:recur3.1}
q(x)  =  \frac{ \{ [x^2-(a-b)^2][(a+b+1)^2-x^2][x^2-4\sigma^2]
\}^{1/2} }{x(4x^2-1)^{1/2} }
\end{equation}
\begin{equation}\label{eq:recur3.2}
q_0(x) = \frac{2\sigma[a(a+1) - b(b+1)]}{x(x+1)} - 2\delta~.
\end{equation}

This recurrence can be regarded as a dual of the previous one: it
is a symmetric eigenvalue equation with the allowed $2\delta$ as
eigenvalues, and can be related to the dual Hahn polynomials
\cite{avcf1995,fcv.03}. These two three-term recurrence equations
can be unified in a single ``five-term" (or better two-variable
three-term) relationship similar to one introduced by us in Ref.
\cite{lncs1.2013} for the $6j$ symbols.

The recurrence relations can be solved either as an
eigenvalue/eigenvector problem or as a linear algebra problem. In
each case the sign (phase) of the normalized $2j$ symbol must be
set.  For this purpose we have used the following convention: the
sign of $U_{ab(\frac{a+b}{2})}(x,\frac{a-b}{2})$ is $(-1)^{2a}$
and the sign of $U_{ab\sigma}(a+b,\delta)$ is
$(-1)^{a-b-2\sigma}$.

\section{Basic equations for caustics and ridges}\label{s5}

The caustics and the ridges are curves which we can represent on
the screen to establish the asymptotic behavior, and in particular
the quantum-classical boundaries.

From Eq. (\ref{eq:recur1}), defining as usual in semiclassical
approaches \cite{varsh,schgora,schgorb}, $J_1 = a+\frac{1}{2}$,
$J_2 = b+\frac{1}{2}$, $J_3 = x+\frac{1}{2}$, we have an
``oriented area"
\begin{eqnarray}
 S^2 &=& -\frac{1}{16}\left|
 \begin{matrix}
   0              & J_1^2-\alpha^2        & J_2^2-\beta^2        & 1\cr
   J_1^2-\alpha^2 & 0                     & J_3^2-(\alpha+\beta)^2 & 1\cr
   J_2^2-\beta^2 & J_3^2-(\alpha+\beta)^2 & 0                  & 1\cr
   1           & 1                 & 1                  & 0
 \end{matrix}
 \right|\nonumber\\
     &=&F^2+\frac{(\sigma^2-\delta^2)\,J_3^2}{4} -\frac{\sigma\left[(\sigma+\delta)\,J_2^2+(\sigma-\delta)\,J_1^2\right]}{2}~,
\end{eqnarray}
where
\begin{eqnarray}
 F =\frac{\sqrt{(J_1+J_2+J_3)(-J_1+J_2+J_3)(J_1-J_2+J_3)(J_1+J_2-J_3)}}{4}
\end{eqnarray}
is the Archimedes-Heron formula for the area of the triangle
having sides $J_1$, $J_2$, and $J_3$. According to previous
sections, $m_1 = \sigma + \delta$ and $m_2 = \sigma - \delta$.
Caustics are obtained by imposing $S=0$ (the solution is given
below, Eq.\ref{cau}). Ridges $\delta^*(J_3)$ and $J_3^*(\delta)$
are found both following $\delta$ at fixed $J_3$
\begin{equation}
 \delta^*(J_3) = \sigma\frac{J_1^2-J_2^2}{J_3^2}.
\end{equation}
or viceversa following $J_3$ at fixed $\delta$
\begin{equation}
 J_3^*(\delta) = \sqrt{J_1^2+J_2^2 + 2(\sigma^2 - \delta^2)}~.
\end{equation}
The upper and lower caustics are then conveniently expressed
explicitly, as a function of $J_3$, as follows:
\begin{equation}\label{cau}
 \delta_{\pm}(J_3) = \delta^*(J_3) \pm 2\,F\frac{\sqrt{J_3^2-4\,\sigma^2}}{J_3^2}~.
\end{equation}
Differentiating the latter equation, one finds the cases when
caustics exhibit a cusp: this occurs when $\sigma=\pm(J_1-J_2)/2$,
namely for a $3j$ invariant

 with respect to Regge symmetry, the
cusp will occur either in the lower or upper left corner of the
screen, according to the sign, as shown in the next section.

\section{Images}\label{s6}

\begin{figure}
  \center
  \subfigure[][$\sigma=0$]{
    \includegraphics[width=.3\textwidth, clip, trim= .7cm .7cm 1.7cm 1.7cm, page=1]{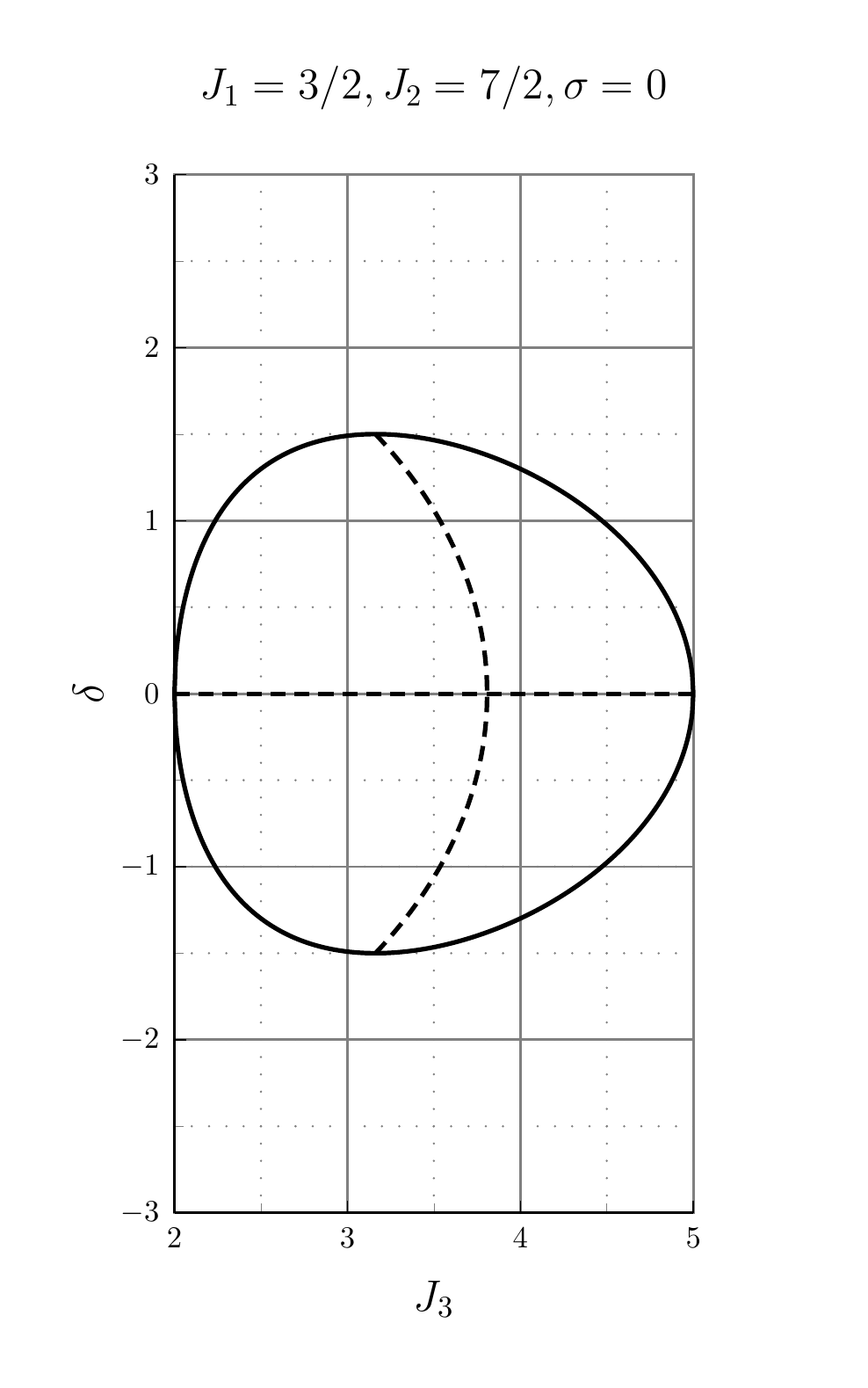}
  }
  \subfigure[][$\sigma=\pm 1/2$]{
    \includegraphics[width=.3\textwidth, clip, trim= .7cm .7cm 1.7cm 1.7cm, page=2]{FIG.pdf}
  }
  \subfigure[][$\sigma=\pm 1$]{
    \includegraphics[width=.3\textwidth, clip, trim= .7cm .7cm 1.7cm 1.7cm, page=3]{FIG.pdf}
  }\\
  \subfigure[][$\sigma=\pm 3/2$]{
    \includegraphics[width=.3\textwidth, clip, trim= .7cm .7cm 1.7cm 1.7cm, page=4]{FIG.pdf}
  }
  \subfigure[][$\sigma=\pm 2$]{
    \includegraphics[width=.3\textwidth, clip, trim= .7cm .7cm 1.7cm 1.7cm, page=5]{FIG.pdf}
  }
  \subfigure[][$\sigma=\pm 5/2$]{
    \includegraphics[width=.3\textwidth, clip, trim= .7cm .7cm 1.7cm 1.7cm, page=6]{FIG.pdf}
  }
  \caption{\label{FIG1}Caustic and ridge plots (continuous and dashed curves, respectively) of $3j$ symbols for $J_1=3/2, J_2=7/2$ and for the allowed values of $\sigma$.}
\end{figure}

\begin{figure}
  \center
  \includegraphics[width=.8\textwidth]{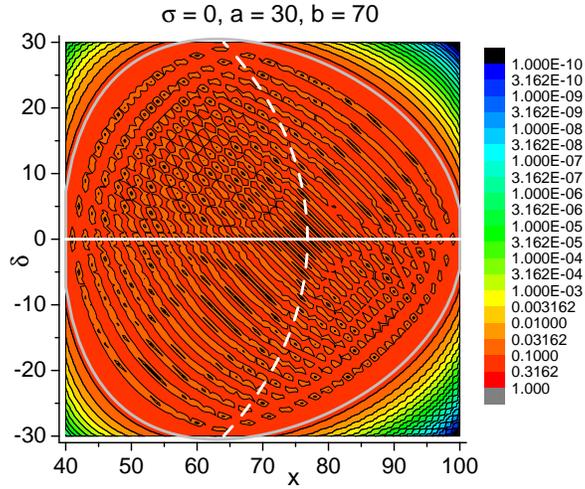}
  \caption{\label{FIG2a}The gray loop is the caustic line, and the dashed and solid white lines are
           the ridges. The color map log scale plots are for the absolute value of the $3j$ coefficients,
           and the range is $10^{-10}$ to $1$.}
\end{figure}

\begin{figure}
  \center
  \includegraphics[width=.8\textwidth]{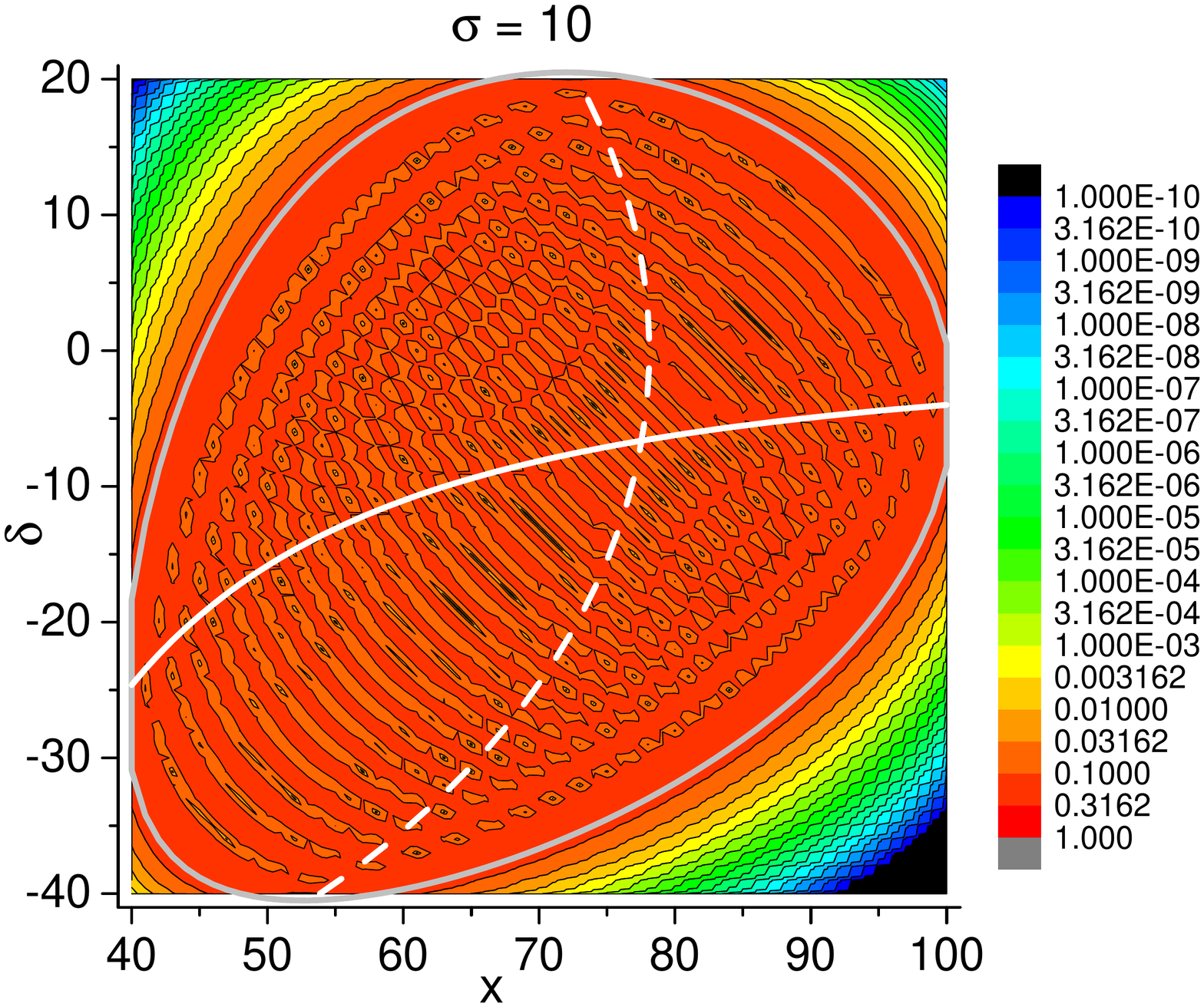}
  \caption{\label{FIG2b}As in Fig. \ref{FIG2a} for $\sigma=10$.}
\end{figure}

\begin{figure}
  \center
  \includegraphics[width=.8\textwidth]{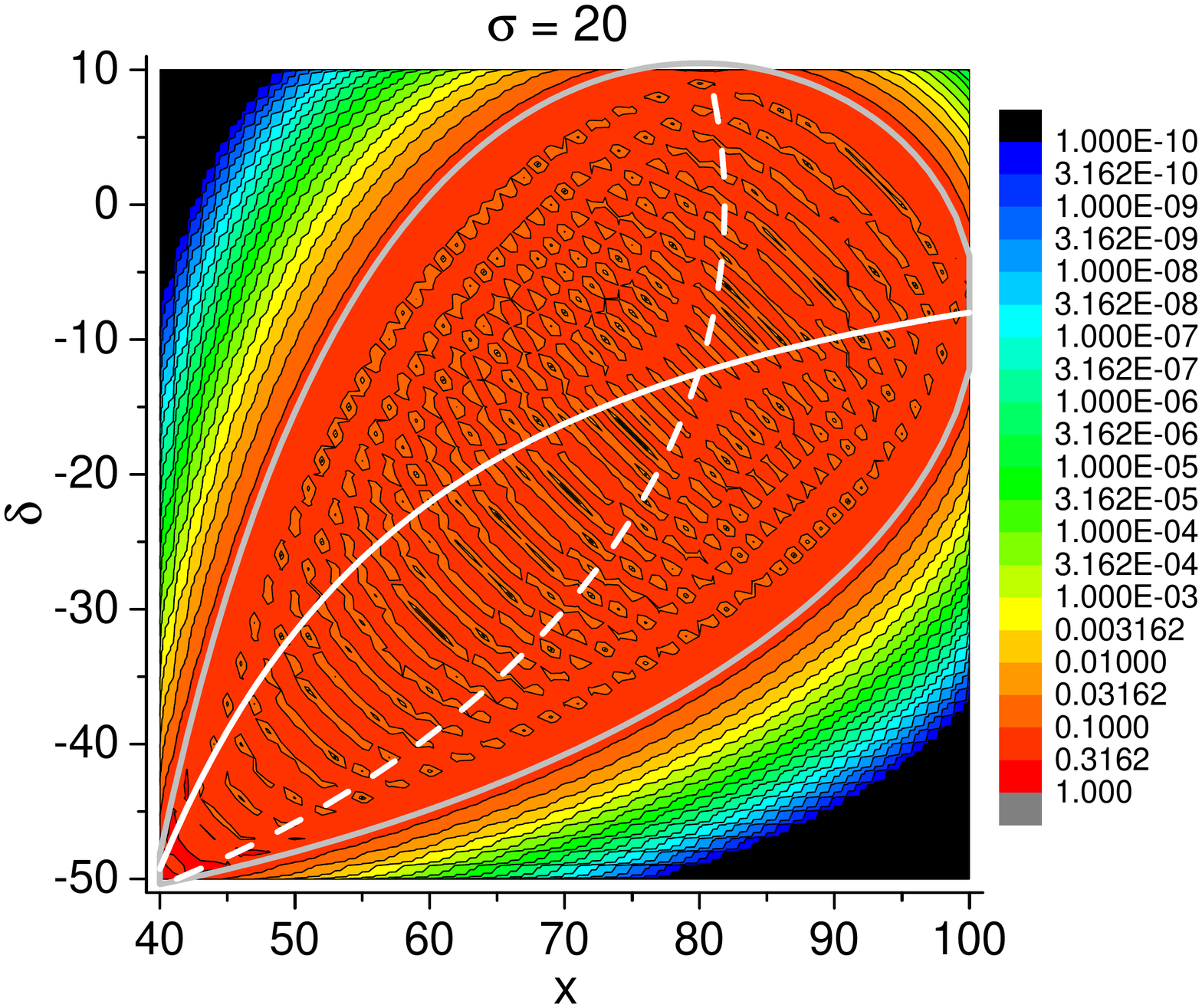}
  \caption{\label{FIG2c}As in Fig. \ref{FIG2a} for $\sigma=20$.}
\end{figure}

\begin{figure}
  \center
  \includegraphics[width=.8\textwidth]{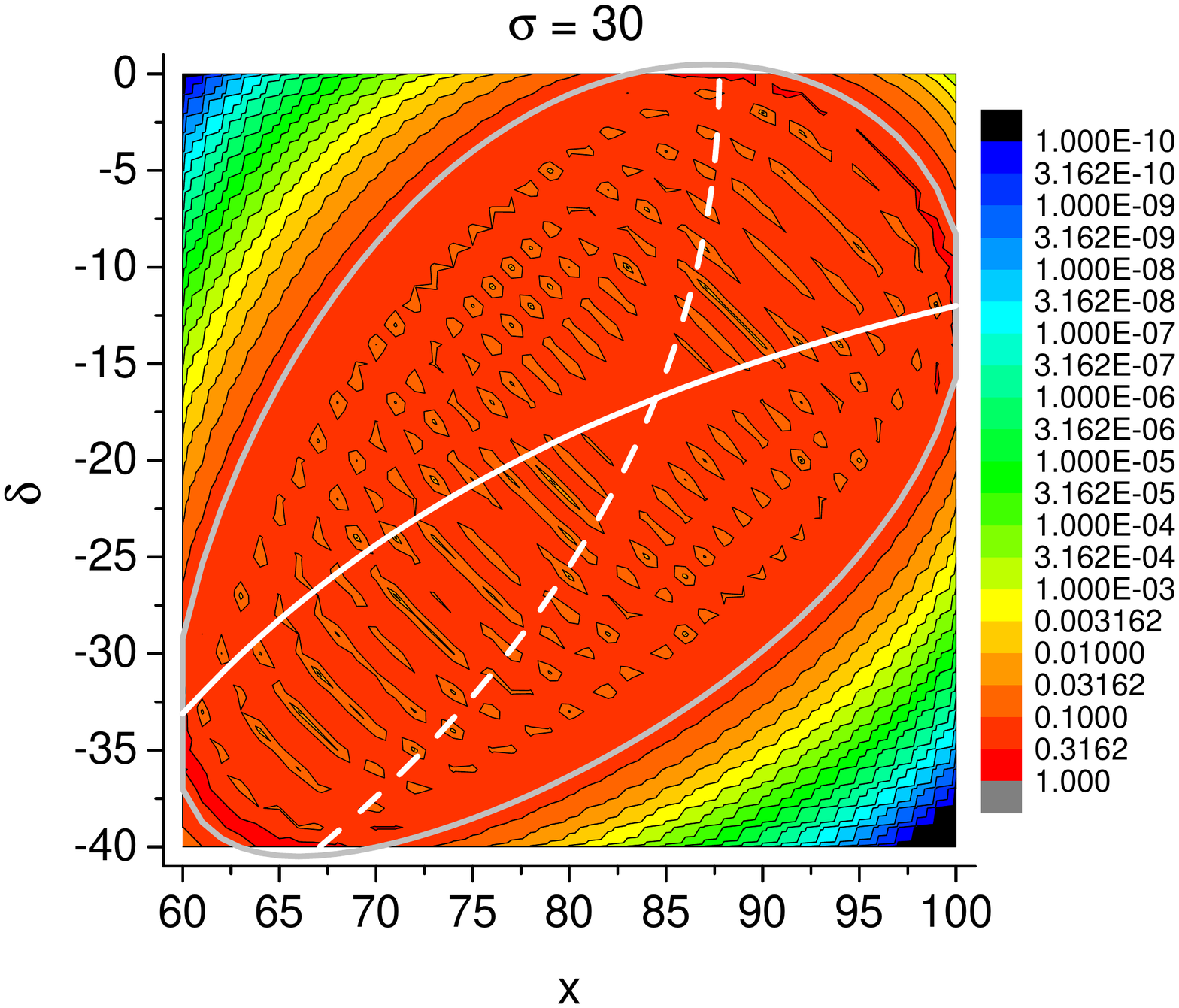}
  \caption{\label{FIG2d}As in Fig. \ref{FIG2a} for $\sigma=30$.}
\end{figure}

\begin{figure}
  \center
  \includegraphics[width=.8\textwidth]{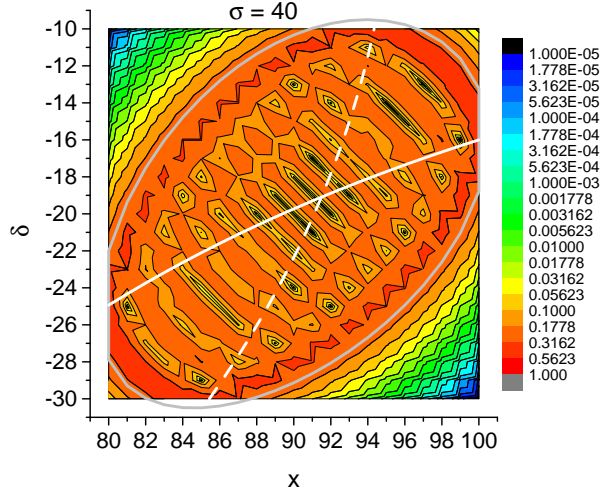}
  \caption{\label{FIG2e}As in Fig. \ref{FIG2a} for $\sigma=40$, but with a range from $10^{-5}$ to $1$.}
\end{figure}

\begin{figure}
  \center
  \includegraphics[width=\textwidth, clip, trim= .7cm .7cm 1.7cm 1.7cm, page=7]{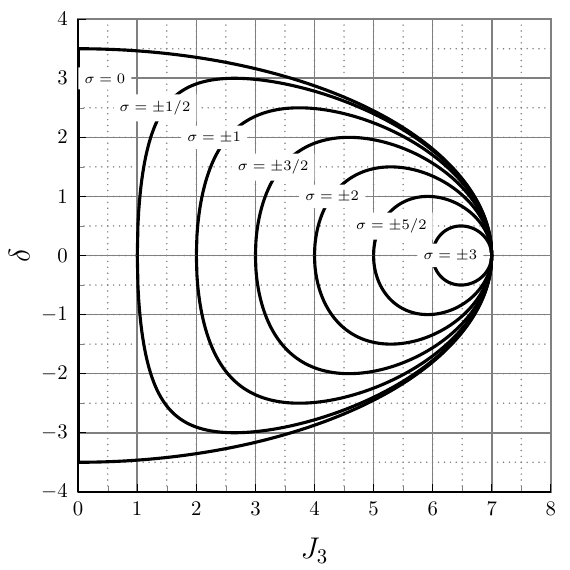}
  \caption{\label{FIG3}Plots of caustic for $3j$ symbols for $J_1=J_2=7/2$ and for allowed values of $\sigma$.}
\end{figure}

The paper concludes with illustrations of the above treatment
(Figs. \ref{FIG1} - \ref{FIG3}), showing results of exact
calculations of $3j$ symbols, accompanied by drawings of the
asymptotic (semiclassical) behavior.
The analysis of the phenomenology is carried out guided by \cite{ponzano1968semiclassical,doi:10.1021/jp905212a,aaf.08,roberts1999classical,rbfaal.10}.
The square screens have $x$ in abscissas and $\delta$ in ordinates.

\section{Concluding remarks}\label{s7}
The study of these coefficients is important as orthogonal basis
sets in discretization algorithms
\cite{fcv.03,aquilanti:3792,ac.00}. In fact, not considered here
are their limits to spherical and hyperspherical harmonics
(\emph{e.g.} d matrix) when entries are large. This makes them
useful for expanding continuous functions on grids.

An important topics is the semiclassical dynamics associated to
the $3j$ symbols, that can be worked out similarly to that for
$6j$'s. The interesting geometrical interpretations
\cite{lncs2.2014,ac.01,mr0194029,springerlink:10.1007/s00214-009-0519-y,jpa2012,nikiforovsuslovuvarov199110,jeanmarclevyleblond1965symmetrical,bilo9,doi:10.1021/jp905212a,ragni2010}
are also currently being investigated.

\section{Appendix: Permutational and ``mirror'' symmetries}\label{app}

\begin{figure}
  \center
  \includegraphics[width=\textwidth, clip, trim= .1cm .1cm .1cm .1cm, page=9]{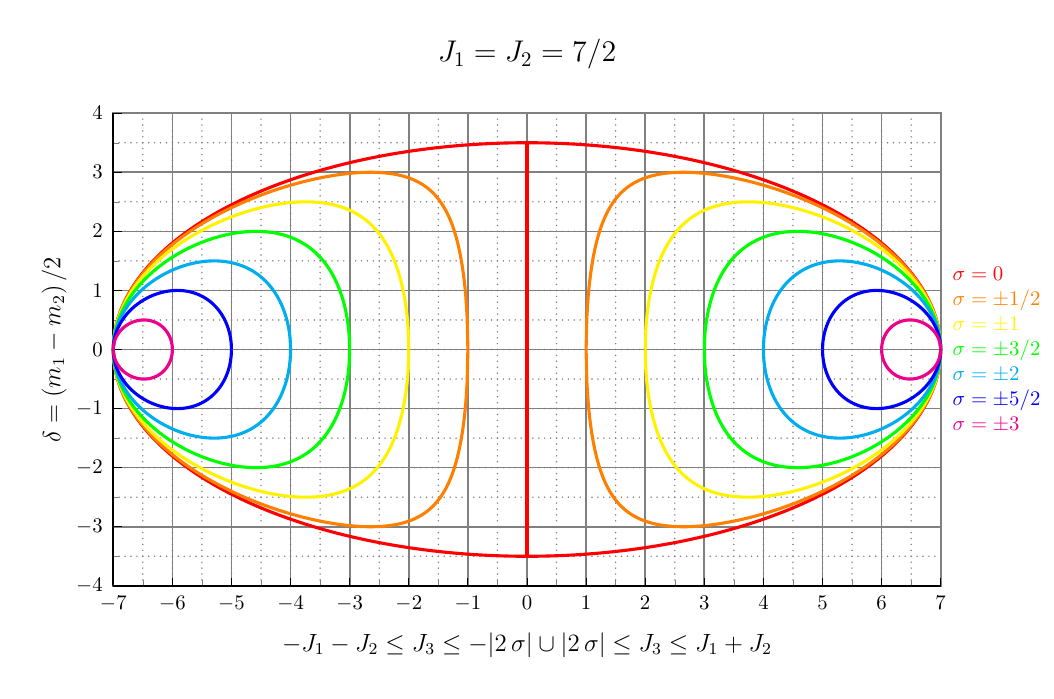}
  \caption{\label{FIG4}Caustic plots of $3j$ symbols for $J_1=J_2=7/2$ and for allowed values of $\sigma$.}
\end{figure}

Symbols related by exchange of a column involve a phase change,
\emph{e.g.}, in our notation
\begin{equation}\label{app1}
\itrej{a}{b}{x}{\alpha}{\beta}{\gamma} =
(-1)^{a+b+x}\itrej{b}{a}{x}{\beta}{\alpha}{\gamma}~.
\end{equation}
Similarly, changing signs for all projections
\begin{equation}\label{app2}
\itrej{a}{b}{x}{\alpha}{\beta}{\gamma} =
(-1)^{a+b+x}\itrej{a}{b}{x}{-\alpha}{-\beta}{-\gamma}~.
\end{equation}
Substituting our variables, given in Eq. (\ref{eq:recur1}), we have
\begin{equation}\label{app3}
\itrej{a}{b}{x}{\sigma+\delta}{\sigma-\delta}{-2\sigma} =
(-1)^{a+b+x}\itrej{a}{b}{x}{-\sigma-\delta}{-\sigma+\delta}{2\sigma}~,
\end{equation}
which is used in the screen representation of Sec. \ref{s6}.
The mirror symmetry, \emph{i.e.} $a\rightarrow -a-1$,
$b\rightarrow -b-1$, $x\rightarrow -x-1$ permits the
introduction of negative entries, \emph{e.g.}
\begin{equation}\label{app1}
\itrej{a}{b}{x}{\alpha}{\beta}{\gamma} = (-1)^{b-x-a}\itrej{a}{b}{-x-1}{\alpha}{\beta}{\gamma}
\end{equation}
as illustrated in Fig. \ref{FIG4}.

\bibliographystyle{splncs}


\end{document}